# Observation of angle-dependent mode conversion and mode hopping in 2D annular antidot lattice

Nikita Porwal[1], Anulekha De[2], Sucheta Mondal[2], Koustuv Dutta[2], Samiran Choudhury[2], Jaivardhan Sinha[2], Anjan Barman[2] & P. K. Datta[1]

We report spin-wave excitations in annular antidot lattice fabricated from 15 nm-thin $Ni_{80}Fe_{20}$ film. The nanodots of 170 nm diameters are embedded in the 350 nm (diameter) antidot lattice to form the annular antidot lattice, which is arranged in a square lattice with edge-to-edge separation of 120 nm. A strong anisotropy in the spin-wave modes are observed with the change in orientation angle ($\phi$) of the in-plane bias magnetic field by using Time-resolved Magneto-optic Kerr microscope. A flattened four-fold rotational symmetry, mode hopping and mode conversion leading to mode quenching for three prominent spin-wave modes are observed in this lattice with the variation of the bias field orientation. Micromagnetic simulations enable us to successfully reproduce the measured evolution of frequencies with the orientation of bias magnetic field, as well as to identify the spatial profiles of the modes. The magnetostatic field analysis, suggest the existence of magnetostatic coupling between the dot and antidot in annular antidot sample. Further local excitations of some selective spin-wave modes using numerical simulations showed the anisotropic spin-wave propagation through the lattice.

Advancements in lithography techniques have triggered fabrication of artificially patterned magnetic nanostructures, which have potential applications in high density magnetic storage[1], memory[2], logic[3], transistor[4], while strongly coupled magnetic nanowires, nanodots and magnetic antidot lattices have potential applications in on-chip data communication and processing devices. The latter, known as magnonic crystal (MC), the magnetic analogue of electronic, photonic and phononic crystals, have strongly emerged during last one decade and extensive research have been carried out on ferromagnetic nanowire arrays[5], dot arrays[6–10] and antidot arrays[11] to that end. Interesting results such as resonant mode splitting[12], mode crossover[7,13], mode conversion[14], mode localization[15], mode softening[12,16], configurational anisotropy[13,17], magnonic mini band formation[18] have been observed and new device concepts such as magnonic filters[19], phase shifters[20], splitters[21], coupler[22], waveguides[23], logic devices[24], transistors[4] etc. have been demonstrated. New devices with greater functionalities and sustainabilities would require more robust design and control parameters, and hence, newer and more complex structures have been continuously introduced by the researchers such as bicomponent magnonic crystals[25–28], magnonic quasicrystals[29] and three-dimensional magnonic crystals[30]. However, the improved functionalities often come at the expense of more complicated fabrication processes, such as multistep lithography and two-photon photolithography. Binary magnonic crystal (BMC)[31–35] is another class of crystal, which can be formed using simpler lithography processes by placing two structures of same material next to each other forming the basis of the crystal, and it can also provide more control parameters for tuning the magnonic bands. Very few works have been reported thus far in the binary structures. In 2008, N. Singh et al.[31] has reported the fabrication and direct mapping of magnetization states of binary magnonic crystals calling as anti-ring samples. Later, Ding et al. studied extensively the static and dynamic behaviour of anti-ring samples using broadband FMR, MOKE and MFM[33]. The potential application of the structure in biosensing was subsequently proposed and demonstrated by Sushruth et al.[34]. In a BMC, having two antidot sublattices alternating-diameters create non-uniform demagnetizing fields. This significantly alters the magnetization reversal than that of single antidot lattice and hence shows field-amplitude-dependent spin-wave (SW) mode transformation[36–38]. In a diatomic nanodot array a distinct change of SW mode characteristics within the diatomic unit and array from that of single nanodots for different bias field orientations, indicating strong magnetostatic interaction among the dots have been observed[35]. More recently, we introduced BMC in the form of an annular antidot lattice[32]. We have shown the bias-field-dependent resonant modes and mode profiles and observed some new interacting modes in such structure using time-resolved magneto-optical Kerr effect microscope (TR-MOKE) and micromagnetic simulations.

[1]Department of Physics, Indian Institute of Technology Kharagpur, Kharagpur, WB, 721302, India. [2]Department of Condensed Matter Physics and Material Sciences, S. N. Bose National Centre for Basic Sciences, Block JD, Sector III, Salt Lake, Kolkata, 700 106, India. Correspondence and requests for materials should be addressed to P.K.D. (email: pkdatta@phy.iitkgp.ac.in)

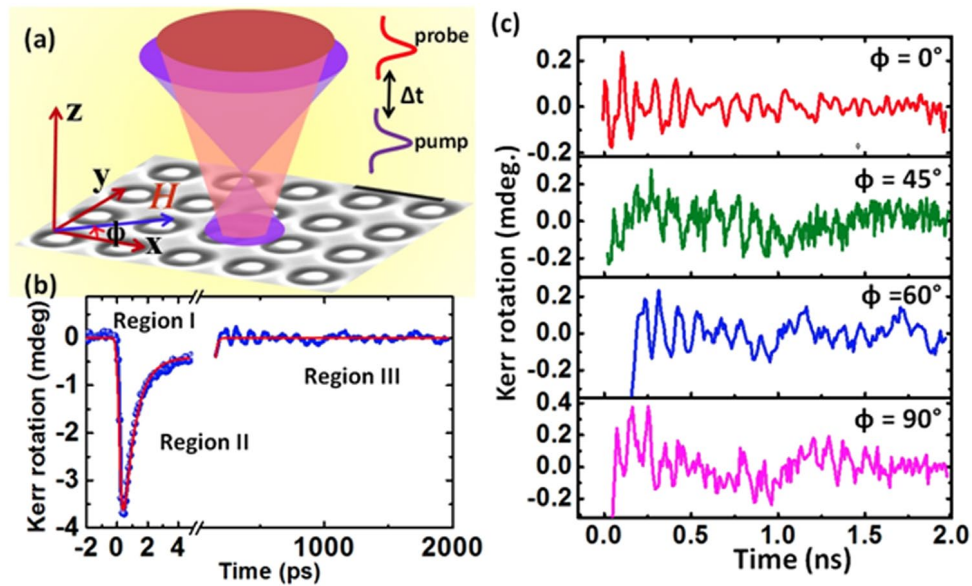

**Figure 1.** (**a**) The experimental geometry with the bias magnetic field (*H*) applied in-plane of the array with a slight out-of-plane tilt. (**b**) Time-resolved Kerr rotation data showing three different regions for ϕ = 60°. Solid lines are fits. (**c**) Experimental background subtracted time-resolved Kerr rotation data for some specific orientations of the bias magnetic field. The corresponding in-plane angles of the bias field are shown in the figures.

Here, we investigate how the SW dynamics of annular antidot lattice is modified by rotating the in-plane direction of the bias magnetic field by using TR-MOKE. Due to the change of interaction between the dot and antidot at different orientations of bias magnetic field, we observe an interesting feature of mode hopping for some selective modes. Furthermore, rotational anisotropy of certain modes is also observed. The origin of these features has been explained by comparing them with individual dot lattice (DL) and antidot lattice (ADL) modes. Identification of the spatial profiles of the modes by micromagnetic simulations reveals the nature of change of interaction with in-plane bias magnetic field angles. We believe that our findings could be important in the contest of the present efforts to understand the SW dynamics in the emerging field of binary magnonic crystals.

## Results and Discussion

The annular antidot lattice is a two-dimensional (2D) binary magnonic crystal in the form of embedded nanodots in a periodic $Ni_{80}Fe_{20}$ antidot lattice arranged in square symmetry. The sample is fabricated on e-beam evaporated permalloy ($Ni_{80}Fe_{20}$, Py hereon) film using focused ion beam lithography. The details of the sample fabrication are described elsewhere[32]. The thickness of the Py film is 15 nm. The scanning electron micrograph (SEM) of the sample is shown in Fig. 1(a). The diameter, D of the antidot is 360.0 ± 3.5 nm, while the dot has diameter (*d*) of 170.0 ± 1.5 nm. The lattice constant is 480 ± 3 nm. The atomic force microscope (AFM) and magnetic force microscope (MFM) images of the sample are presented in ref.[32]. The ultrafast magnetization dynamics of the sample is measured by using an all-optical TR-MOKE microscope in polar Kerr geometry (see Fig. 1(a)). An in-plane bias magnetic field is applied at an angle of ~15° to the plane of the sample. The magnetization is aligned uniformly along the direction of the applied magnetic field. An all-optical pump beam is used to excite the magnetization dynamics of the sample and a time-delayed probe laser beam is used to probe the evolution of magnetization as a function of delay time with respect to pump excitation. Details of the set-up is described elsewhere[39]. The experiment is performed at room temperature and under ambient condition.

To validate the experimental results, we have also performed micromagnetic simulations using object oriented micromagnetic framework (OOMMF)[40]. The mask used in the simulations is derived from the SEM image using 2D periodic boundary conditions with a cell size of 5 nm × 5 nm × 15 nm. The simulation is carried out in a square lattice with 308 × 308 cells in an area of 1.9 μm × 1.9 μm. The material parameters are extracted by measuring the variation of precessional frequency (*f*) with bias magnetic field (*H*) for a Py thin film and by fitting the result using Kittel formula[41]. For Py, the parameters are gyromagnetic ratio $\gamma = 17.6\,MHz\,Oe^{-1}$, anisotropy field $H_k = 0$, saturation magnetization $M_s = 860\,emu\,cm^{-3}$, and exchange stiffness constant $A = 1.3 \times 10^{-6}$ erg $cm^{-1}$. The exchange stiffness constant *A* is obtained from literature[42]. These parameters are consistent with our previous work[32]. To excite the precessional dynamics, a pulsed magnetic field of peak amplitude of 30 Oe, rise/fall time of 10 ps and pulse duration of 50 ps is applied perpendicular to the sample plane under a constant bias field of $H = 1.08$ kOe. The damping coefficient $\alpha = 0.008$ is used during dynamic simulations, which is standard for Py. The spatial distribution of power and phase of different SW modes are mapped by using a home-built code Dotmag[43]. The magnetostatic field distribution is calculated from LLG micromagnetic simulator[44].

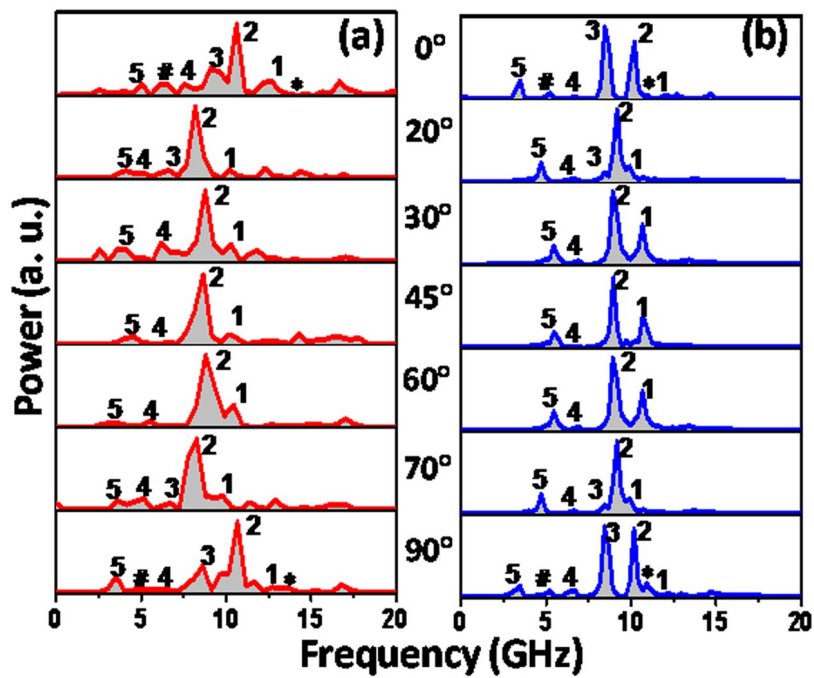

**Figure 2.** (**a**) Power vs frequency spectra of the sample at those specific orientations of bias magnetic field (ϕ) obtained from using (**a**) time-resolved MOKE experiment and (**b**) micromagnetic simulations.

The experimental geometry is depicted in Fig. 1(a). The time-resolved Kerr rotation data shown in Fig. 1(b) consist of three temporal regions. Region I corresponds to the ultrafast demagnetization followed by fast re-magnetization (region II). The precessional dynamics (region III) appears as an oscillatory signal on top of the slow re-magnetization of the time-resolved Kerr rotation data. We have performed precise measurement of time-resolved Kerr rotation and fitted the data with three temperature model using analytical expression given in ref. [45]. From the fitting we obtained the ultrafast demagnetization time and fast relaxation as 130 fs and 780 fs respectively, this is followed by slower relaxation as 13 ps. A bi-exponential background is subtracted from the precessional data, and fast Fourier transform (FFT) is performed to obtain the SW spectra. The background subtracted Kerr rotation data for $\phi = 0°$, 45°, 60° and 90° are shown in Fig. 1(c). Spin-wave spectra are obtained by performing the FFT of the time-resolved Kerr rotation for $0° \leq \phi \leq 180°$. Figure 2(a) shows SW spectra only up to 90° for some specific angles commensurate with the symmetry of the lattice. We obtain rich SW spectra with clear modulation with ϕ. Five SW modes are identified in the spectra which show periodic variation with ϕ whereas the '*' and '#' marked peaks are present only at $\phi = 0°$ and 90°. Our earlier work[32] showed that these two modes are present at all bias-field values at $\phi = 0°$ although their intensity increases at low field strength. However, those modes disappear at an intermediate angular range and hence, we discard these modes while analysing the angular variation of SW frequencies. The frequency of mode 1 decreases systematically between $0° \leq \phi \leq 20°$, following which it becomes nearly constant for $20° \leq \phi \leq 70°$ and then increases for $70° \leq \phi \leq 90°$. Mode 2 is generally the highest intensity mode for the full angular range. Its angular variation is significantly different from that of mode 1. Its frequency remains at a constant level for $0° \leq \phi \leq 15°$, followed by a sharp reduction to another constant level for $20° \leq \phi \leq 70°$, which then increases to a constant level for $75° \leq \phi \leq 105°$. Mode 3 shows starkly different behaviour. This mode exists only for specific angular ranges such as $0° \leq \phi \leq 20°$ and $70° \leq \phi \leq 90°$ and vanishes in between. Here, we mainly focus only on these three modes as shown in Fig. 2. Modes 4 and 5 appear with very low power, which is sometime buried under noise. Furthermore, their angular variation shows distorted four-fold rotational symmetry, as shown in the Supplementary Material. The simulated SW spectra for annular antidot lattice with varying ϕ are shown in the Fig. 2(b), which qualitatively reproduce all the important features of the experimental modes. The variation in the frequency and power of modes 1 and 3 with ϕ is more prominent in simulated frequency spectra as opposed to the experimental spectra. In order to get further insight into the evolution of modes 1 to 3, we have performed detailed simulations as shown in Supplementary Materials. There, we observe that the frequencies of all three modes periodically vary as described already. The power of these modes shows interesting behaviour for $10° \leq \phi \leq 30°$. As ϕ increases power of mode 1 gradually increases up to 30°, while power of mode 2 increases up-to 20° followed by a gradual decrease up to 30°. On the other hand, mode 3 loses its power and completely disappears at 22°.

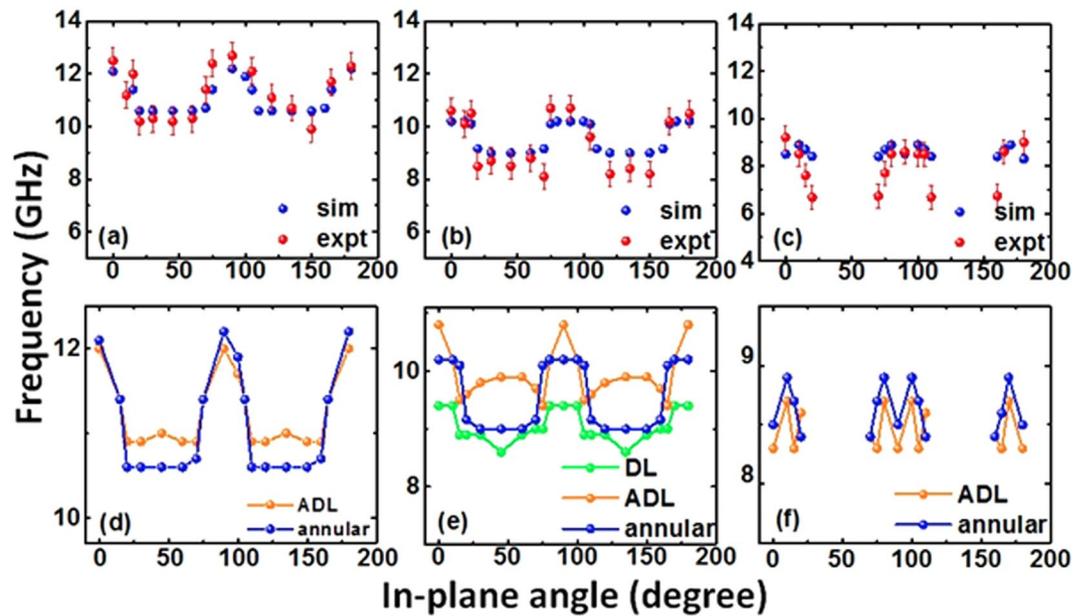

**Figure 3.** Variation of SW frequency of (**a**) mode 1 (**b**) mode 2 and (**c**) mode 3 of the annular antidot lattice with the in-plane angle of the bias field varying from 0° to 180° at a fixed $H = 1080$ Oe obtained from both experiment and simulations. The origin of (**d**) mode 1 (**e**) mode 2 and (**f**) mode 3 of annular antidot lattice studied by comparing with individual DL and ADL.

The quantitative discrepancy between the simulated and experimental SW spectra is attributed to the imperfections such as edge roughness and deformation of the real annular antidot lattice, modifying the internal magnetic field and the stray field experienced by precessing spins in the close proximity of the hole edges. In simulations, we have assumed hole edges that are almost perfect, i.e., the edges are rough only on the lateral scale of the implemented cell size. The variation of experimental and simulated SW frequencies with in-plane bias field angle for modes 1, 2 and 3 of the annular antidot lattice are shown in Fig. 3(a–c). Mode 1 shows distorted four-fold symmetry with flattened valley where we observe that its frequency gets locked in $20° \leq \phi \leq 70°$ regime. In contrast to this, mode 2 hops between two definite frequency levels in the angular range of $0° \leq \phi \leq 15°$ and $20° \leq \phi \leq 70°$, which is repeated periodically. Mode 3 appears and disappears periodically at angles of $0° \leq \phi \leq 20°$ and $20° < \phi \leq 70°$. A close inspection in the simulation results of mode 3 shows a small double peak-like variation, which is repeated with $\phi$ but that could not be reproduced in the experiment due to noise.

To understand the origin of these modes, we have further simulated individual dot lattice (DL) and antidot lattice (ADL) of same dimensions and compared the results with the annular antidot lattice (AAL). As observed from Fig. 3(d), the highest frequency mode (mode 1) of the AAL is originated from the ADL with slight modification of frequency values. However, mode 2 (see Fig. 3(e)) exhibits a clear anisotropic behaviour in ADL for $20° \leq \phi \leq 70°$. Its frequency falls sharply from 0° to 20° followed by a gradual increase up to 45° and another gradual decrease from 45° to 70°. From 70° to 90° the frequency increases sharply back to the value observed at 0°. Mode 2 of the DL shows a distorted four-fold symmetry. Consequently, the variation of mode 2 of AAL stems from the superposition of the angular variation of this mode in DL and ADL. Mode 3 (Fig. 3(f)) of the annular antidot lattice originates from ADL which shows identical behaviour for this mode in AAL.

Figure 4 shows the simulated power and phase maps of the resonant modes calculated using a home-built code Dotmag[43]. The SW mode profiles are of two types: modes which are quantized in both DL and ADL, or modes that are extended in ADL but uniformly distributed in the DL. The nature of the modes depends on both the orientation of $H$ as well as the dipolar interaction in the system. At $\phi = 0°$, mode 1 is a standing SW mode in backward volume (BV) geometry with quantization number $n = 5$ within the dots and $n' = 7$ within the antidots of the AAL, although the power within the dot is negligible. Mode 2 is a centre mode of the dots but in antidots it shows quantized mode with $n' = 3$. Mode 3 is an extended mode in the vertical channels of the antidots, i.e. in the Damon-Eshbach (DE) geometry without having any significant power in the dots. As $\phi$ changes up to 45°, mode 1 retains its quantized behaviour with the quantization axis rotating in commensuration with the bias field direction. For mode 2, the angular variation also retains the qualitative nature of the modes but the mode axes rotate in commensuration with the bias field direction. The power is mainly concentrated in the dots. Mode 3, however show a mode conversion with the bias field orientation, transforming from an extended mode to a localized mode at the antidot edge.

To understand the mode flattening and mode hopping of modes 1 and 2, respectively, we locally excited the dynamics using a sine function containing the frequency of the resonant mode at the region shown in Fig. 5 by the grey bar of $2380 \times 60$ nm² area. We observe that the SW corresponding to mode 1 propagates to a large distance along the vertical direction with significant power (about 25 dB) at $\phi = 0°$. With increasing $\phi$, the propagation gradually decreases and stops propagating at 45°. However, mode 2 suddenly transforms from a propagating mode to a localized mode as ϕ is increased from 0° to 20° and retains its character up to 45°. This anisotropic nature of SW propagation with bias field orientation is the reason behind the observed flattening and hopping of the mode 1 and mode 2.

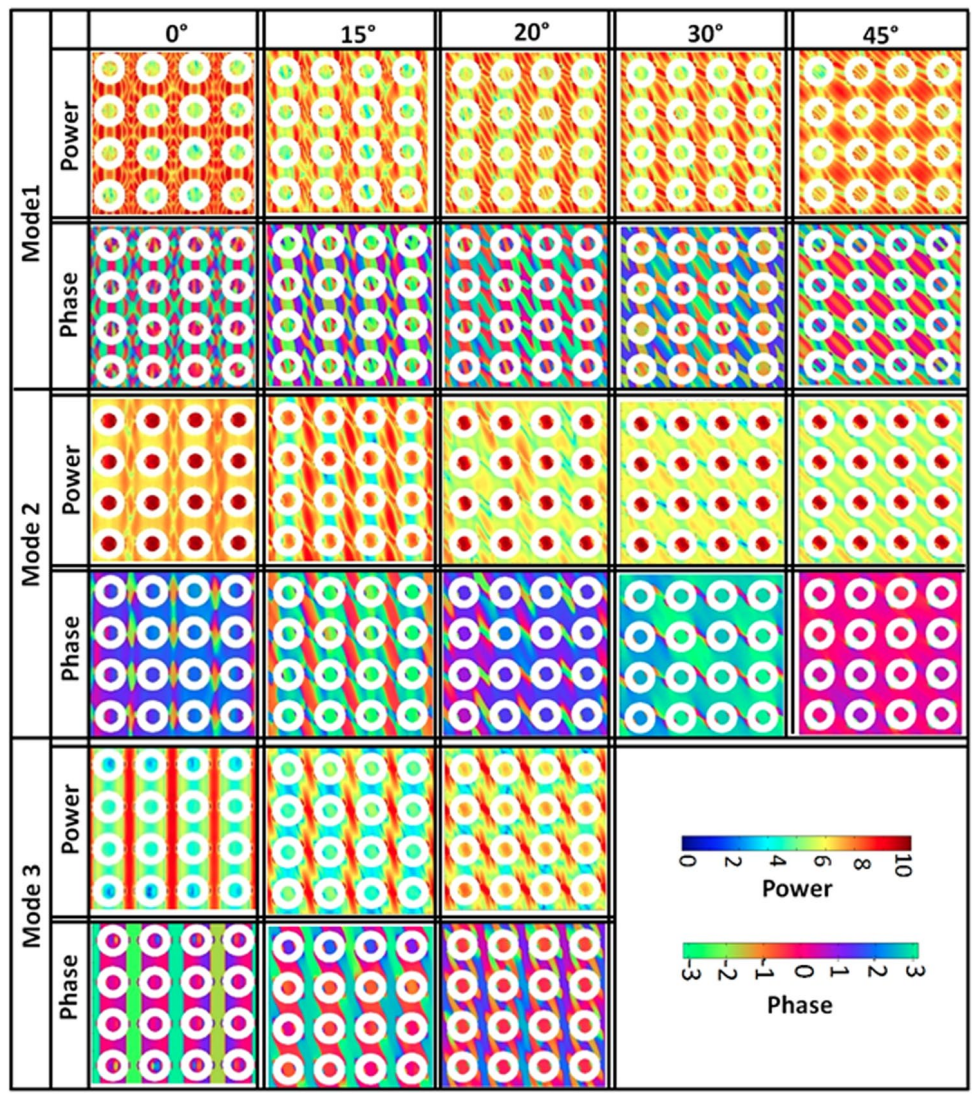

**Figure 4.** Spin-wave mode profile of the sample for different orientations of the in-plane bias magnetic field. The colour map used for the mode profiles is shown in the figure.

To further understand the dynamics, we have numerically calculated[44] the magnetostatic field distributions in the AAL and the corresponding contour plots are shown in Fig. 6 (a) for four different orientations (ϕ) of the bias magnetic field. We took the line scan of the x-component of magnetostatic field ($B_x$) which shows local maxima and minima across the three regions, named region A (within the dot), region B (air) and region C (antidot channel) with varying ϕ. One can readily observe that the interaction between the magnetostatic field lines increases at the region B and region C with increase in ϕ value, which is probably due to the increase of the interactions at the edges of dots and antidot regions. This is also clear from the linescans of the magnetostatic field as shown in Fig. 6(b).

With the increase in ϕ, the internal field decreases in all three regions as shown in Table 1. The variation in $B_x$ is maximum in region C.

The decrease in the internal field with increasing the bias field orientation is probably responsible for variation in the frequencies of the observed SW modes. For deeper understanding of mode hopping, we have simulated the magnetostatic field of the AAL, ADL and DL with varying ϕ between 0° and 180° and compared the total internal magnetic field ($B_a$) with the frequency variation of mode 2 as shown in Fig. 7. We observe that the variation of $B_a$ in AAL, DL and ADL show qualitatively similar behaviour to the frequency variation of mode 2, giving rise to its observed anisotropy, while for the other modes the situation is far more complex due to the complicated interaction field profiles.

Since, the bias field dependent SW spectra showed some significant variation in the spectra with bias field, it is important to understand the angular variation of the SW modes at a lower bias field. To this end, we have simulated the SW spectra at $H = 620$ Oe, and the detailed results are presented in the Supplementary Material. In this case all the seven modes show distinct angular dispersion as opposed to the $H = 1080$ Oe, where only five modes show angular dispersion. Moreover, while some of the modes (mode 1 and mode 2) retain their angular dispersion behaviour, some other modes show slight or significant change in the angular dispersion behaviour. This is most likely related to the variation of the internal field profile, which gets more nonuniform with the reduction of the bias field. This is interesting as the two very important phenomena like mode hopping and mode flattening (locking), are stable with the bias field magnitude. On the other hand, the other modes become more sensitive to angular variation with the reduction of bias field, implying their non-reliability for magnonic device applications. Nevertheless, they also give opportunity to tune the modes more efficiently.

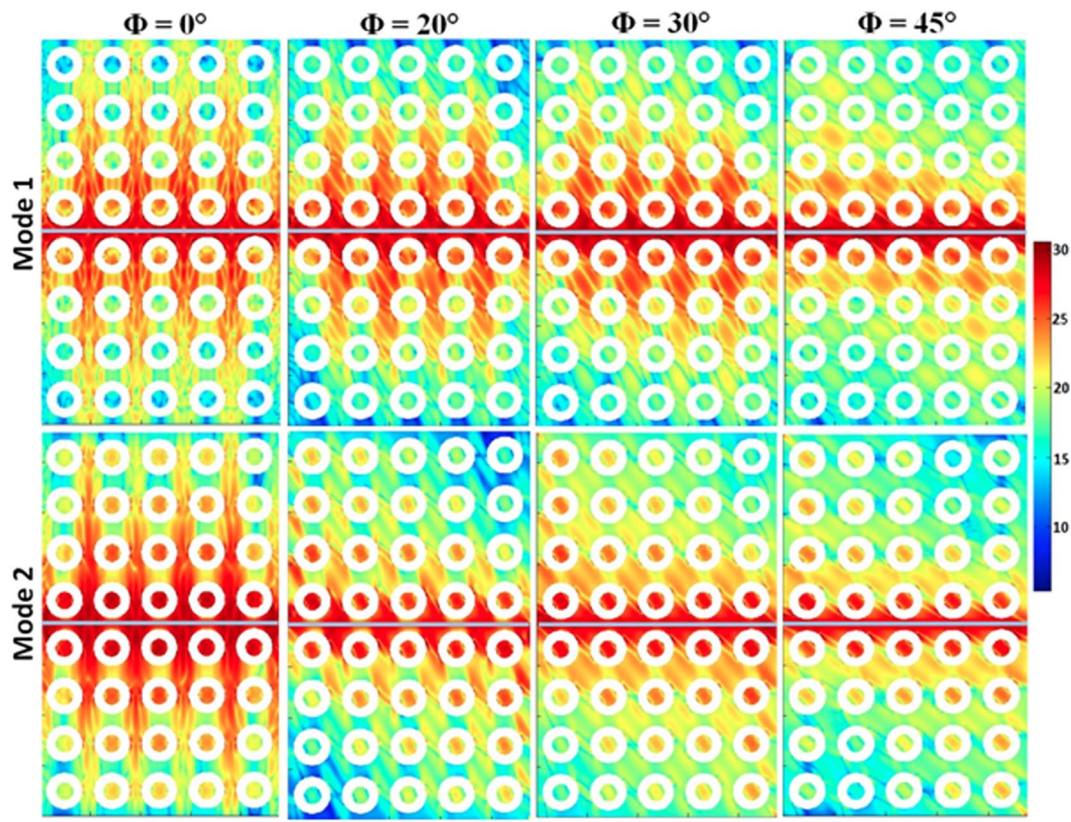

**Figure 5.** Simulated power profiles of spin-wave mode 1 and mode 2 excited locally over the rectangular strip place at middle of the sample for φ = 0°, φ = 20°, φ = 30°, and φ = 45° geometries.

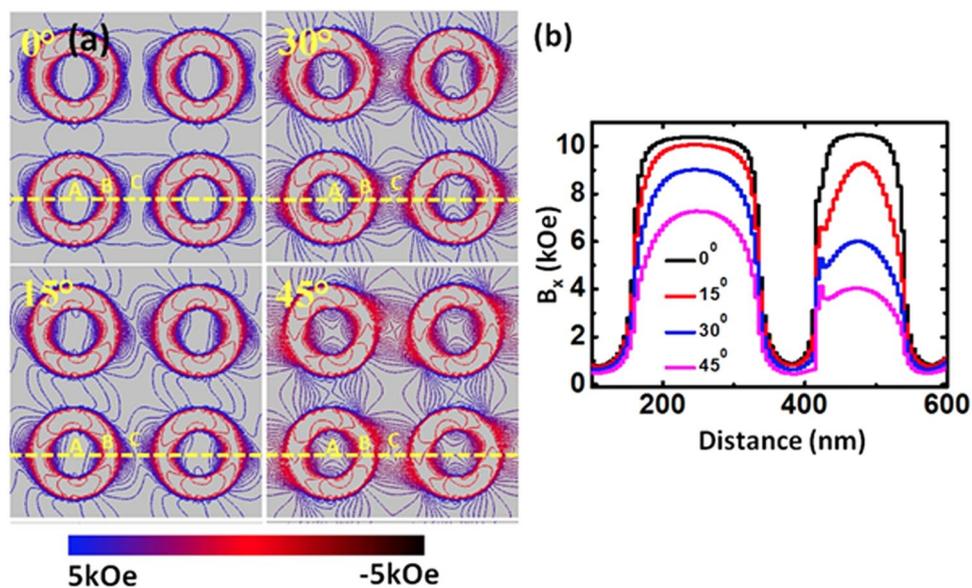

**Figure 6.** (**a**) Contour maps of simulated magnetostatic field distributions for different orientation of the in-plane bias magnetic field of the sample. The strength of the stray field is represented by the color map at the bottom of the figure. (**b**) Comparison of the simulated magnetostatic field components $B_x$ for specific φ taken along the dotted yellow lines from the sample.

| Angular variation | Magnetostatic field (kOe) | | |
|---|---|---|---|
| | Region A | Region B | Region C |
| 0° | 10.42 | 0.89 | 10.48 |
| 45° | 7.24 | 0.43 | 4.05 |

**Table 1.** Effect of variation of bias field orientation on the magnetostatic fields of three different regions of the AAL.

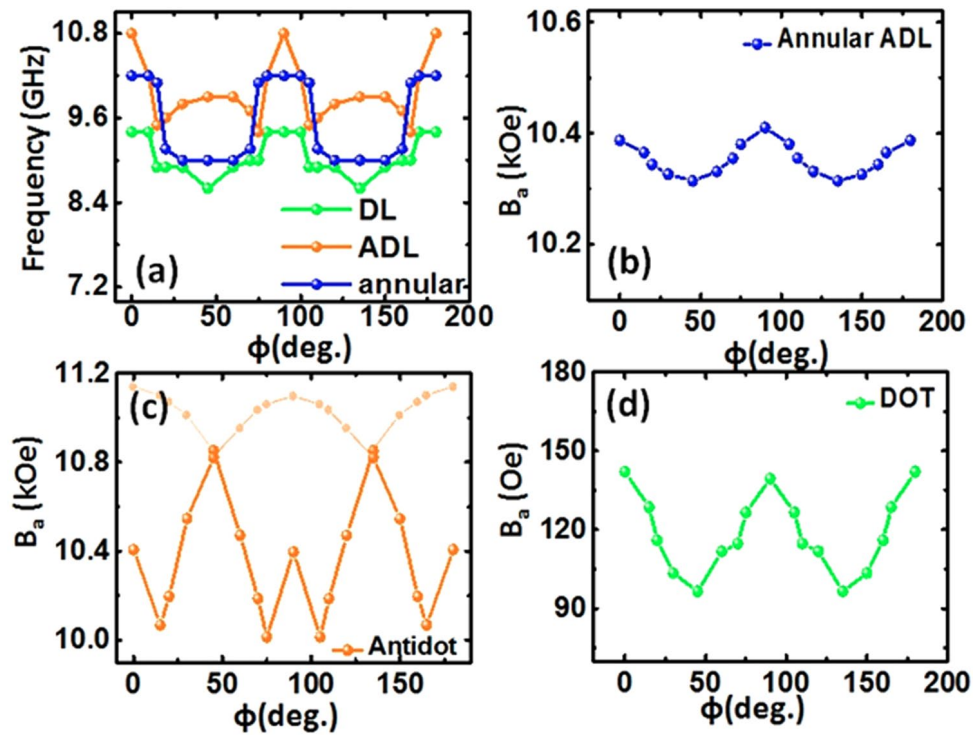

**Figure 7.** (**a**) Variation of SW frequency of mode 2 of the annular antidot lattice, only DL and only ADL with the in-plane angle of the bias field varying from 0° to 180° at a fixed $H = 1080$ Oe for comparison. The internal magnetic field of annular antidot lattice (**b**), only antidot (**c**) and only dot (**d**) with the in-plane angle of the bias field varying from 0° to 180° at a fixed $H = 1080$ Oe.

## Conclusion

Here, we report an experimental and numerical study of the variation of spin-wave dynamics of annular antidot sample with the orientation of the bias magnetic field using TR-MOKE microscope. We observe a flattened four-fold rotational symmetry, mode hopping and mode conversion leading to mode quenching for three prominent spin-wave modes in this lattice with the variation of the bias field orientations. By detailed micromagnetic simulations of spin-wave power and phase maps and magnetostatic field distributions of the annular antidot lattice and the constituent dot and antidot lattices, we have underpinned the origin of the observed mode anisotropies. We have further locally excited the selective modes using numerical simulations and showed the anisotropic spin wave propagation through the lattice, indicating their possible applications in spin-wave filter and other nonlinear spin-wave devices.

## Methods

**Fabrication.** A Py thin film of 15 nm thickness is deposited by using electron beam evaporation on top of Si substrate using an ultrahigh vacuum chamber at a base pressure of $\sim 2 \times 10^{-8}$ torr. The film is immediately transferred to the sputtering chamber for the deposition of capping layer (SiO$_2$) of 5 nm to avoid degradation from natural oxidation during experiment. Sputter deposition is done by rf sputtering at a base pressure of $\sim 2 \times 10^{-7}$ torr and Ar pressure of $\sim 5 \times 10^{-3}$ torr at 60 W of Ar power at a frequency of 13.56 MHz. The annular antidot is patterned on this blanket Py thin film by using liquid Ga$^+$ ion beam lithography (Auriga-Zeiss FIB-SEM microscopes) over a square area of 10μm$^2$. The optimal values of voltage and current for milling are found to be 30 keV and 5 pA respectively. At this beam current, we get sufficient etch rate yet limit the spot size to be around 50 nm. The thickness of the Py film is smaller than the stopping rate of Ga$^+$ ion beam at 30 keV, which ensures that the ions stop within the Si layer underneath the Py (15 nm). We verified the etching depth by using atomic force microscopy (AMF) measurement, which show etching depth of ~25 nm.

**Measurement.** A two-colour optical pump-probe set-up is used to measure ultrafast magnetization dynamics of the sample. The second harmonic ($\lambda = 400$ nm, pulse width = 100 fs, fluence = 10 mJ/cm$^2$) of a mode locked Ti-sapphire laser (Tsunami, Spectra Physics) is used to pump the sample, whereas the time-delayed fundamental beam ($\lambda = 800$ nm, pulse width = 80 fs, fluence = 2 mJ/cm$^2$) is used to probe the magnetization dynamics by measuring the Kerr rotation as a function of time-delay between pump and probe beams. The pump and probe beams are made collinear before incidence on the sample using the same microscope objective (N.A. 0.65) focused to spot sizes of 1 μm and 800 nm, respectively. A magnetic field is applied at an angle of about 15° from the sample-plane to have a finite demagnetizing field along the direction of the pump beam. The pump beam modulates this out-of-plane demagnetizing field component to induce precession of magnetization in the sample. The in-plane component of this field is the bias magnetic field ($H = 1.08$ kOe), whose orientation ($\phi$) is varied to study the angular variation of SW dynamics of this sample (as shown in Fig. 1(a)). The sample is rotated between $0° \leq \phi \leq 180°$ using a high precision rotary stage while keeping the microscope objective and magnetic field constant. The pump and the probe beams are carefully placed on the same region of the array for each value of $\phi$. The experimental time window of 2 ns is found to be sufficient for extracting the important information of the magnetization dynamics. The experiment is performed at room temperature and under ambient condition.

### Acknowledgements
P.K.D. acknowledges the SGDRI (UPM) project of IIT Kharagpur for support. A.B. acknowledges S. N. Bose National Centre for Basic Sciences for financial support (Project No. SNB/AB/18–19/211). N.P. acknowledges IIT Kharagpur for senior research fellowship. A.D., S.M. and K.D. acknowledge DST for INSPIRE fellowship, J.S. acknowledges DST for Ramanujan fellowship, while S.C. acknowledges S. N. Bose National Centre for Basic Sciences for senior research fellowship.

### Author Contributions
N.P. and P.K.D. planned the project. J.S. and S.C. prepared Py thin film. N.P. fabricated annular antidot sample. S.M., A.D., K.D and N.P. performed Kerr Microscopy experiment. N.P. led the investigation and prepared the first draft. N.P. performed Micromagnetic simulations. P. K.D., A.B. provided supervision and contributed to the discussion. N.P., S.M, A. D., P.K.D and A.B. modified the draft after consultation with all co-authors.

### Additional Information
**Supplementary information** accompanies this paper at https://doi.org/10.1038/s41598-019-48565-8.

**Competing Interests:** The authors declare no competing interests.

**Publisher's note:** Springer Nature remains neutral with regard to jurisdictional claims in published maps and institutional affiliations.